\documentclass[journal=jacsat,manuscript=article]{achemso}

\usepackage[version=3]{mhchem} % Formula subscripts using \ce{}
\usepackage[T1]{fontenc}
\newcommand{\alscn}{Al\textsubscript{1-x}Sc\textsubscript{x}N}
\newcommand{\alscnAA}{Al\textsubscript{0.73}Sc\textsubscript{0.27}N}
\newcommand{\albn}{Al\textsubscript{1-x}B\textsubscript{x}N}
\newcommand{\alyn}{Al\textsubscript{1-x}Y\textsubscript{x}N}
\newcommand{\gascn}{Ga\textsubscript{1-x}Sc\textsubscript{x}N}
\author{Md Redwanul Islam}
\affiliation[Kiel University]
{Department of Materials Science, Kiel University, Kiel, Germany}
\email{mdis@tf.uni-kiel.de}
\author{Niklas Wolff}
\affiliation[Kiel University]
{Department of Materials Science, Kiel University, Kiel, Germany}
\author{Tom-Niklas Kreutzer}
\affiliation{Fraunhofer Institute for Silicon Technology (ISIT), Itzehoe, Germany}
\author{Georg Schönweger}
\affiliation[Kiel University]
{Department of Materials Science, Kiel University, Kiel, Germany}
\alsoaffiliation{Fraunhofer Institute for Silicon Technology (ISIT), Itzehoe, Germany}
\author{Margaret Brown}
\affiliation{Department of Metallurgical and Materials Engineering,
Colorado School of Mines, Golden, Colorado 80401, USA}
\author{Maike Gremmel}
\affiliation[Kiel University]
{Department of Materials Science, Kiel University, Kiel, Germany}
\author{Patrik Stra\v{n}\'{a}k}
\affiliation{Fraunhofer Institute for Applied Solid State Physics IAF, Freiburg, Germany}
\author{Lutz Kirste}
\affiliation{Fraunhofer Institute for Applied Solid State Physics IAF, Freiburg, Germany}
\author{Geoff L. Brennecka}
\affiliation{Department of Metallurgical and Materials Engineering,
Colorado School of Mines, Golden, Colorado 80401, USA}
\author{Simon Fichtner}
\email{sif@tf.uni-kiel.de}
\affiliation[Kiel University]
{Department of Materials Science, Kiel University, Kiel, Germany}
\alsoaffiliation{Fraunhofer Institute for Silicon Technology (ISIT), Itzehoe, Germany}
\author{Lorenz Kienle}
\email{lk@tf.uni-kiel.de}
\affiliation[Kiel University]
{Department of Materials Science, Kiel University, Kiel, Germany}

\title[An \textsf{achemso} demo]
  {Improved Leakage Currents and Polarity Control through Oxygen Incorporation in Ferroelectric \alscnAA\ Thin Films}


\begin{document}

%%%%%%%%%%%%%%%%%%%%%%%%%%%%%%%%%%%%%%%%%%%%%%%%%%%%%%%%%%%%%%%%%%%%%
%% The abstract environment will automatically gobble the contents
%% if an abstract is not used by the target journal.
%%%%%%%%%%%%%%%%%%%%%%%%%%%%%%%%%%%%%%%%%%%%%%%%%%%%%%%%%%%%%%%%%%%%%
\begin{abstract}
  This article examines systematic oxygen (\textbf{O})-incorporation to reduce total leakage currents in sputtered wurtzite-type ferroelectric \alscnAA\ thin films, along with its impact on the material structure and the polarity of the as-grown films. The \textbf{O} in the bulk \alscnAA\ was introduced through an external gas source during the reactive sputter process. In comparison to samples without doping, \textbf{O}-doped films showed almost a fourfold reduction of the leakage current near the coercive field. In addition, doping resulted in the reduction of the steady-state leakage currents by roughly one order of magnitude sub-coercive fields. Microstructure analysis using X-ray diffraction and scanning transmission electron microscopy (STEM) revealed no significant structural degradation of the bulk \alscnAA. In case of the maximum \textbf{O}-doped film, the \textit{c}-axis out-of-plane texture increased by only 20\% from 1.8\textdegree\ and chemical mapping revealed a uniform distribution of oxygen incorporation into the bulk. Our results further demonstrate the ability to control the as-deposited polarity of \alscnAA\ via the \textbf{O}-concentration, changing from nitrogen- to metal-polar orientation. Thus, this article presents a promising approach to mitigate the leakage current in wurtzite-type \alscnAA\ without incurring any significant structural degradation of the bulk thin film quality, thereby making ferroelectric nitrides more suitable for microelectronic applications.
\end{abstract}

%%%%%%%%%%%%%%%%%%%%%%%%%%%%%%%%%%%%%%%%%%%%%%%%%%%%%%%%%%%%%%%%%%%%%
%% Start the main part of the manuscript here.
%%%%%%%%%%%%%%%%%%%%%%%%%%%%%%%%%%%%%%%%%%%%%%%%%%%%%%%%%%%%%%%%%%%%%
\section{Introduction}
Ferroelectricity in (0001) oriented wurtzite-type (\textit{w-})\alscn\ thin films opened up new pathways for its integration into functional devices. The material has a high out-of-plane polarization ($P_{r}$) up to $\sim$130 $\mu C/cm^{2}$ and a large tunable coercive field ($E_{c}$) ranging from 2-6 MV cm\textsuperscript{-1}.\cite{fichtner2019alscn} A significant aspect of this material is its simplicity of production, as it is the first nitride ferroelectric that integrates with both CMOS and GaN technologies in a straightforward fashion.\cite{fichtner2019alscn, leone2020metal, wang2021fully, kim2023scalable, mikolajick2021next} Ferroelectric \textit{w}-\alscn\ thin-films can also be considered to be free from wake-up effects, have excellent data retention, and show high temperature stability.\cite{gremmel2024interplay, islam2021exceptional, wang2023ferroelectric, drury2022high, alma998135459402341}. Its ability to electrically write stable polarization domain states into the thin-film structure has attracted significant interest for numerous applications, especially in non-volatile memory for neuromorphic computing, high electron mobility transistors (HEMT), microelectromechanical system (MEMS) based sensors, actuators and harsh environment electronics.\cite{kim2023scalable, mikolajick2021next, streicher2023enhanced, krause2022alscn, islam2021exceptional, drury2022high} The emergence of ferroelectricity in \textit{w}-\alscn\ has also resulted in the discovery of other wurtzite-type ferroelectrics, such as \albn, \gascn, and \alyn, which demonstrated comparable stable ferroelectricity along their \textit{c}-axis.\cite{hayden2021ferroelectricity, wang2021fullygascn, wang2023yaln} Ferroelectricity in these materials comes with two accessible polarization states known as metal (M-) and nitrogen (N-) polar.\cite{wang2021fully, wang2023yaln, hayden2021ferroelectricity} Because of this bipolar configuration, the films always exhibit box-shaped polarization hysteresis (\textit{P-E}) characteristics, regardless of the film microstructure (from fiber textured to epitaxial).\cite{islam2023comparative, schonweger2022fully, yang2024emerging} Among all the possible aforementioned deposition techniques, sputtering of \textit{w}-\alscn\ on Si substrates through intermediate metal electrodes (e.g., Pt, Mo) is a suitable method, especially for achieving ferroelectricity with high Sc concentrations ranging from 25 to 43 at. \% as well as low deposition temperature budget from room temperature to 450 \textdegree C.\cite{fichtner2019alscn, tsai2021room, streicher2023enhanced, wolff2023demonstration, wang2020molecular} These films are fiber textured, have a well-defined interface with the electrodes and appear to be free from chemical segregation.\cite{schonweger2023grain, islam2023comparative, wolff2023demonstration}. It is important to recognize that the maximum attainable \textbf{Sc} concentrations through sputtering on the aforementioned Si systems is only limited by the phase transition threshold where the solid-solution transforms from wurtzite to cubic AlScN, causing a complete loss of polarization in the overall thin film.\cite{akiyama2009enhancement, akiyama2009influence} 

\medskip

One of the yet unresolved issues with the ferroelectric nitrides comes with their intrinsically high $E_c$, which is close to the material's breakdown field ($E_c$ $>$ 0.5*$E_{breakdown}$).\cite{fichtner2019alscn, schonweger2022fully} This causes high leakage currents during switching, which is rather independent of the film microstructure.\cite{schonweger2023structural} Dedicated growth methods such as MBE or MOCVD aiming for excellent crystalline quality of \textit{w}-\alscn\ thin-films on GaN showed no significant improvement of the leakage currents compared to sputtered fiber-textured thin films.\cite{wang2021fully, leone2020metal, wolff2023demonstration} Instead, recent investigation indicates that because of the better thermal coefficient matching between the AlScN and Si substrate, sputtered films on Si substrates (with metal Pt bottom electrodes) have marginally lower leakage profile and average $E_c$ than sputtered epitaxial films on GaN. Although these films on Si substrates have higher density of structural defects such as grain boundaries.\cite{islam2023comparative, schonweger2023grain} Thus, an alternative approach is required to deal with the leakage current issue in \textit{w}-\alscn\ thin films. It is known that elemental doping in semiconductors is a standard, cost-effective way to tune its electronic properties.\cite{geng2021improved} In this study, we intentionally introduced \textbf{O} as a dopant in 200 nm thick \textit{w-}\alscnAA\ via reactive sputtering onto stacks of [111]Pt/Ti (seed)/SiO\textsubscript{2}/Si (referred as ``Pt/Si'' in the following). Elemental doping using \textbf{O} already proved to strongly impact the electrical an optical properties of III-nitrides as a deep level impurity, including \textit{w}-\alscn.\cite{zhang2024dependence, chen2024oxygen}. Its small atomic radius (0.073 nm) results e.g., in the ability to eliminate point defects such as \textbf{N}-vacancies.\cite{zhu2020effect, slater1964atomic} 
However, the impact of \textbf{O} as a prevalent invasive element during sputtering of \textit{w}-\alscn\ on the microstructure is still not experimentally investigated in detail. Here, we demonstrate methodical \textbf{O}-doping of \textit{w}-\alscnAA\ and reveal that it can substantially improve the leakage current response while preserving the structural integrity of the bulk film. The successful growth of \textit{w}-\alscnAA\ films with a homogeneously distributed and increasing \textbf{O}-concentration was demonstrated by time-of-flight secondary ion mass spectrometry (ToF-SIMS) and energy dispersive X-ray spectroscopy (EDS). The bulk structure of the material was analyzed with high resolution X-ray diffraction (HRXRD) showing only a limited degradation of the crystalline out-of-plane texture by $\sim$20\%, even at high \textbf{O} doping levels. All doped films showed at least a four-fold improvement of the ferroelectric switching induced leakage current compared to reference samples deposited without additional oxygen. In the low electric field regime, the steady-state leakage current measurements showed roughly an order of magnitude improvement in the most highly \textbf{O}-doped films. Additionally, electrical measurements combined with piezoresponse force microscopy (PFM) showed a transition of as-deposited polarity with increase of \textbf{O}-doping in the films which changed from N- to M-polarity via intermediate mixed polar states. Transmission-mode ultraviolet-to-visible light spectroscopy  (UV-Vis) on a similar set of epitaxial samples on sapphire substrates showed a minor drop in both direct and indirect bandgap values. We speculate these minor drop of bandgap indicates modulation of intrinsic defect states (filling of the \textbf{N}-vacancies) via DX centers. These speculations also supports Theoretical calculations on the the DX-center behavior in AlN, which indicated local structural distortions to stabilize an acceptor state like O$_{\mathrm{N}}^{\mathrm{-1}}$.\cite{Lee2024, gordon2014hybrid} Furthermore scanning transmission electron microscopy (STEM) with elemental mapping showed no chemical segregation in the bulk \textit{w}-\alscnAA\ as well as no trace of $\alpha$-Al\textsubscript{2}O\textsubscript{3} - in either the bulk or at the Pt/\textit{w-}\alscn\ growth interface, whereas both issues were observed before.\cite{yang2014influence} Hence, our results demonstrate the controlled introduction of \textbf{O} atoms inside the bulk \textit{w-}\alscnAA\ during sputter deposition and its overall positive impact on the ferroelectric leakage current density. Furthermore, given the structural similarities between \textit{w-}\alscn\ and other nitride ferroelectrics, we anticipate our reported method of doping might also be widely applicable.\cite{chen2024reactive, moram2008effect, yang2014influence} Therefore, \textbf{O}-doping can enhance the ferroelectric characteristics of \textit{w}-\alscnAA\ thin films, facilitating their application in novel technological domains.

\section{Results and discussion}
The following sections present the determination of \textbf{O}-doping levels during sputter deposition and its effect on the thin-film microstructure and ferroelectric response of \textit{w}-\alscnAA.

\subsection{Determination of O-content in the doped films} 

\begin{figure}[h!]
  \includegraphics[width=0.60\linewidth]{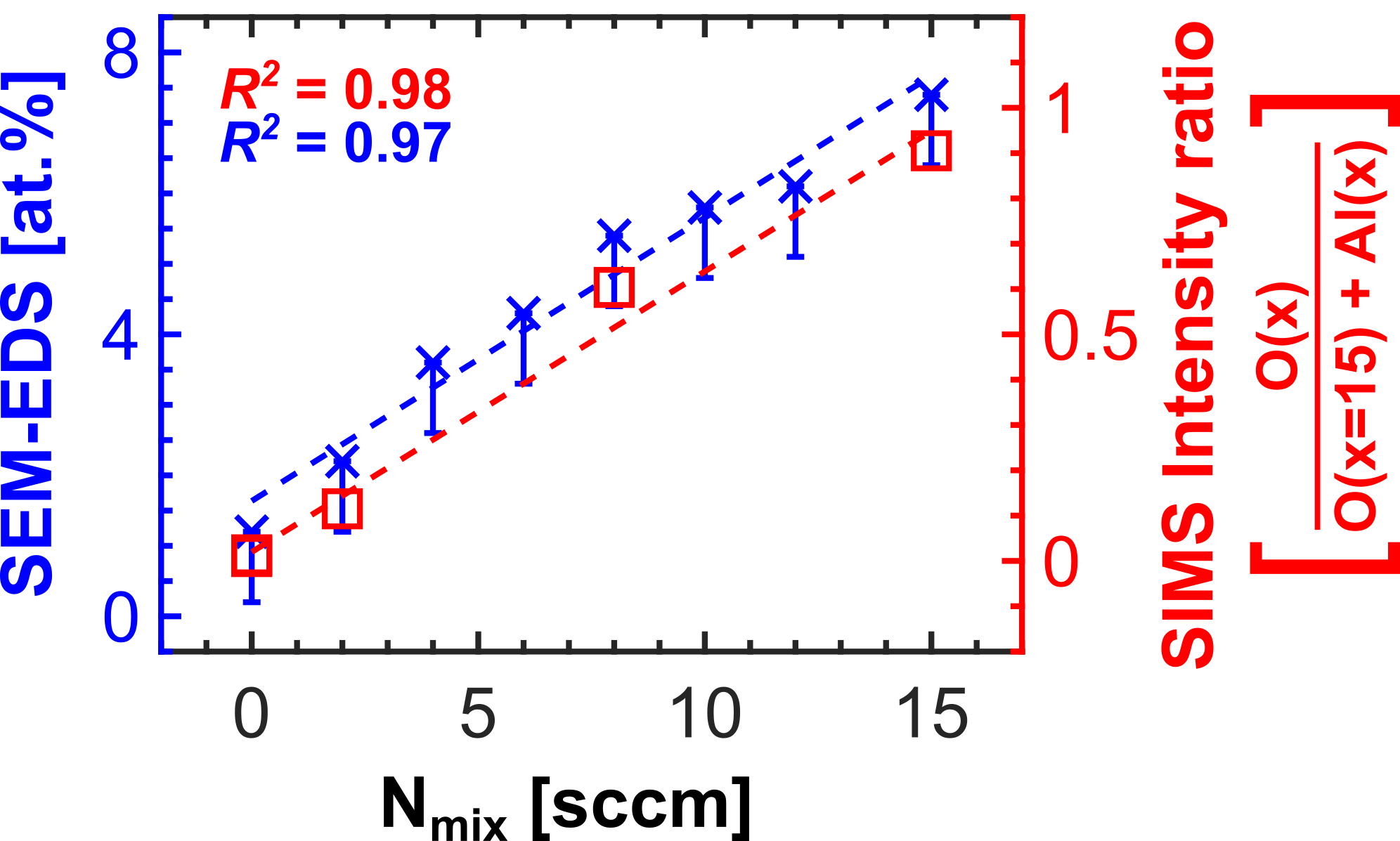}
  \caption{\small{\textbf{O}-content in the bulk \textit{w}-\alscnAA\ quantified from SEM-EDS (left y-axis) and normalized \textbf{O(x)}/(\textbf{O(x=15)} + \textbf{Al(x)}) intensity from negative ion ToF-SIMS measurement (right y-axis). Here, x is the $N_\textrm{mix}$ value of the corresponding specimen and Al(x) is roughly constant for all samples (see Table S1 in the supporting information). The $R^2$ values for linear fitting are listed at the top left of the figure.}}
  \label{fig:oxygencontentOO}
\end{figure}

Systematic \textbf{O}-doping of the bulk \textit{w}-\alscnAA\ was achieved by tuning the overall nitrogen gas flow during deposition from a pure N\textsubscript{2} gas bottle ($N_{\textrm{pure}}$) connected in parallel with another N\textsubscript{2} gas bottle with 2\% O\textsubscript{2} mixture ($N_\textrm{mix}$). The gas flow during the deposition was monitored with a mass flow controller. The \textbf{O}-doping concentration within the bulk was increased by adding $N_\textrm{mix}$ to the overall nitrogen gas flow. For obtaining the optimum film texture, the overall nitrogen flow during deposition was kept constant at 15 sccm. The scenario for doping via external gas flow can be described as following,

\begin{align*}
\overbrace{Ar}^{\textrm{7.5 sccm}}:\ \ \overbrace{N_\textrm{pure}  + N_\textrm{mix}}^{{\textrm{15 sccm}}} = 1:2     
\end{align*} 

The bulk \textbf{O}-concentration of the produced films was measured qualitatively by EDS on a scanning electron microscope (SEM) (all samples) and negative ion mode ToF-SIMS (for samples with $N_\textrm{mix}$ = 0, 2, 8 and 15 sccm) in dependence of the gas mixture. Note, that two distinct $N_\textrm{mix}$ = 0 reference samples were measured in ToF-SIMS to verify the negligible inclusion of \textbf{O} without doping in the bulk \textit{w}-\alscnAA\ thin films. The measurement results are presented in Figure \ref{fig:oxygencontentOO} where the left y-axis shows the approximately linear increase of the \textbf{O} concentration determined from SEM-EDS. The ToF-SIMS measurement are plotted with respect to the right y-axis, where the individual \textbf{O} intensities are normalized by the sum of \textbf{O} intensity at $N_\textrm{mix}$ = 15 and the corresponding \textbf{Al} intensities. This results also demonstrates a comparable linear increasing trend (individual ToF-SIMS measurement profiles are illustrated in Figure S1 of the supporting information). Table 1 in the supporting information summarizes the measured SEM-EDS \textbf{O}-concentrations and ToF-SIMS average \textbf{O} and \textbf{Al} intensities with increasing $N_\textrm{mix}$, noting that this is more of a qualitative statement due to the intrinsically limited accuracies of these chemical analysis methods with respect to the absolute \textbf{O} concentration. Hence, to address the effect of doping, the $N_\textrm{mix}$ values will be used as nomenclature of the samples for the rest of the discussion.
\pagebreak

\subsection{Effects of \textbf{O}-doping on the surface and bulk microstructure}\label{section1: Structural analysis with XRD, SEM, TEM}

\begin{figure}[h!]
  \includegraphics[width=\linewidth]{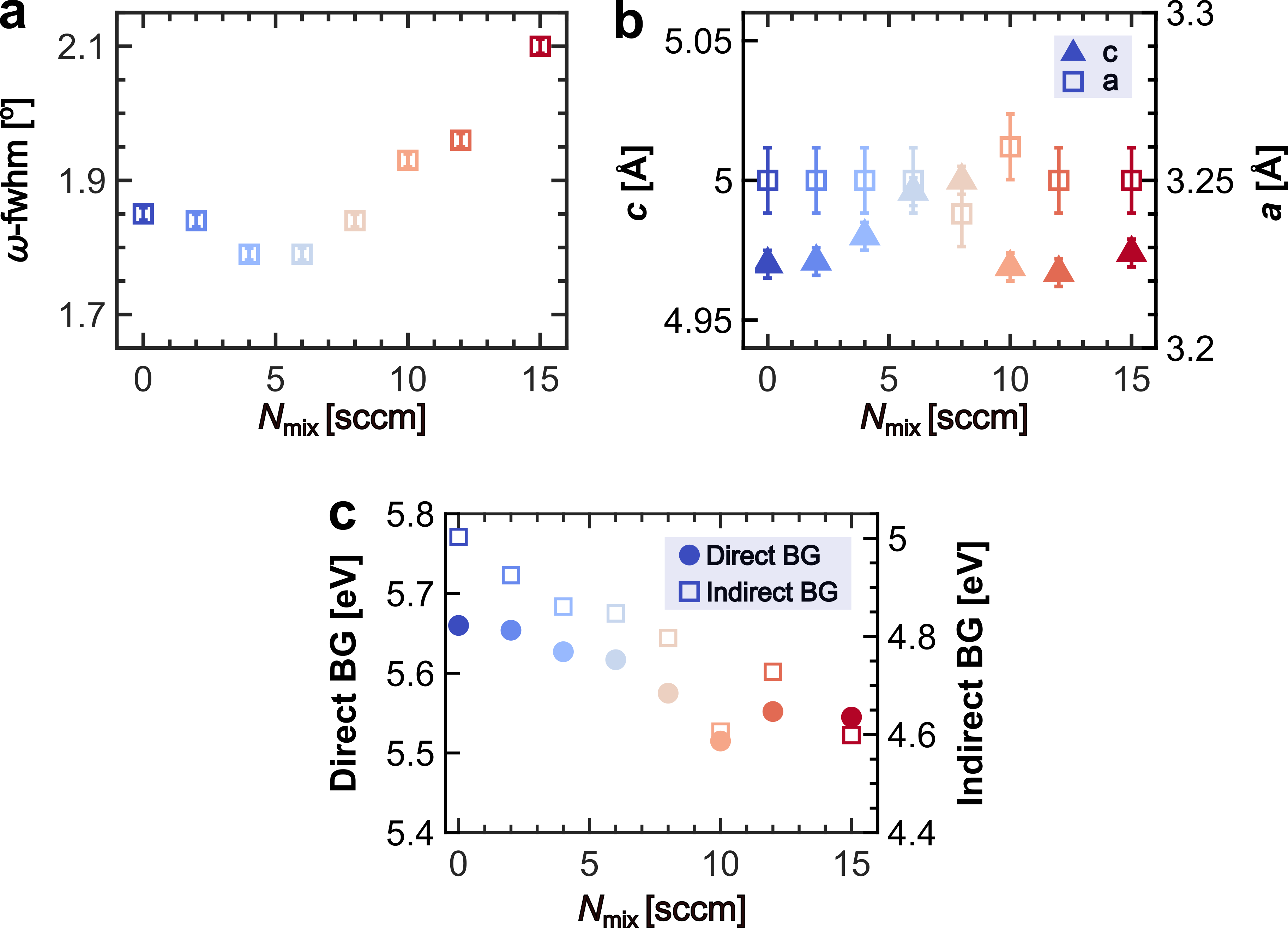}
  \caption{\small{(a) XRD (0002)-rc ($\omega$-FWHM), (b) In and Out of plane \textit{c}-lattice parameters and (c) Estimated direct (circle) and indirect (square) bandgaps of the undoped ($N_\textrm{mix}$ = 0) and \textbf{O}-doped samples ($N_\textrm{mix}$ = 2, 4, 6, 8, 10, 12, 15 sccm).}}
  \label{fig:xrd01}
\end{figure}

The effect of \textbf{O}-doping on the \textit{c}-axis orientation is determined by the 0002 reflection $\omega$-full width at half maximum (FWHM) of high resolution X-ray rocking curve measurements as presented in Figure \ref{fig:xrd01}a. First, all films show clear (0001) texture in $\theta$-2$\theta$ scans (Figure S2, supporting information). Films grown without intended oxygen doping up to $N_\textrm{mix}$ below 8 sccm show no significant changes of the crystallite tilt, with fairly unaffected $\omega$-FWHM around 1.8\textdegree - 1.9\textdegree. However, at higher oxygen input at $N_\textrm{mix}$ $>$ 8 sccm, the $\omega$-FWHM value tends to increase slightly up to $\sim$ 2.1\textdegree\ for $N_\textrm{mix}$ = 15 sccm film. This can be still deemed a good (0001) orientation for \textit{w}-\alscn\ thin films with thicknesses around 200 nm. Hence, the overall increase of columnar grain tilt is below 20\% at a \textbf{O}-doping level as high as $N_\textrm{mix}$ = 15 sccm. Therefore, it can be stated that \textbf{O}-doping cause no heavy degradation of the [0001]-fiber texture of \textit{w}-\alscnAA. Figure \ref{fig:xrd01}b shows the determined \textbf{c} (out-of-plane) and \textbf{a} (in-plane) lattice parameters, where the lattice parameter \textbf{c} shows an incremental increase up to $N_\textrm{mix}$ = 8 sccm and then drops abruptly close to its original value as shown by the filled triangles. In contrast, the \textbf{a} (in-plane) lattice parameter stays constant but shows a small variation around $N_\textrm{mix}$ = 8-10 sccm. Note that the error bar for the \textbf{a} lattice parameter determination via in-plane measurement is larger due to the use of the non-monochromatic beam. It can be assumed that \textbf{O} atoms are likely to replace an \textbf{N} atom (\textbf{O\textsubscript{N}}) in a \textbf{Sc}-coordinated tetrahedra first, since the energy gain when replacing a Sc-N bond with a \textbf{Sc}-\textbf{O} (207.4 kJ mol$^{-1}$) bond is larger compared to the formation of an \textbf{Al}-\textbf{O} bond (133.9 kJ mol$^{-1}$).\cite{luo2007} The bond length of the \textbf{Sc}-\textbf{O} bond, e.g., in cubic Sc\textsubscript{2}O\textsubscript{3} is larger (212 pm) than the bond lengths of Al-N (190 pm, \textit{w}-type AlN) which could result in the observed initial increase of lattice parameter \textit{c}. When the distribution of \textbf{O\textsubscript{N}} atoms inside the material increases it replaces also an increasing number of N-sites in Al-tetrahedra. Dependent on the charge of this \textbf{O\textsubscript{N}}-defect, calculations demonstrated a destabilization of the Al-N bond length for pure AlN leading to structural deformation of the Al-tetrahedra.\cite{gordon2014hybrid, Lee2024} However, calculations of the distortions in \textbf{O\textsubscript{N}} substituted \textbf{Sc}-tetrahedra are missing. It is intriguing to assume that such deformations will cause local strain resulting not only in the observed changes of lattice parameters with increasing oxygen concentration but which could affect the global film stress. 
The evaluation of the thin film stress of doped samples on 6-inch wafers shows a change from tensile to compressive stress with doping and a change in the sign of the slope after passing a similar threshold oxygen concentration reached with 8-10 sccm of $N_\textrm{mix}$ (Figure S3, supporting information). This 
result might indicate that after a certain level of \textbf{O}-doping, the microstructure of \textit{w}-\alscnAA\ thin films undergoes a minor stress relaxation. This small relaxation might be associated as well with the reversal of the as-deposited polar direction as later shown in Figure \ref{fig:polarity01}.
\medskip

Additionally, epitaxially grown samples with similar conditions directly on sapphire were investigated to estimate the optical bandgaps via transmission-mode ultraviolet-to-visible light spectroscopy (UV-Vis). These samples showed similar microstructural trends discussed above as investigated via HRXRD (Figure S4, Supporting information). The use of this separate sample set on sapphire was necessary to conduct as the sample has to be transparent for transmission measurements. Through Tauc analysis of transmission UV-Vis data, estimations of the direct and indirect optical bandgaps were made. As shown in Figure \ref{fig:xrd01}c, we noticed a slight ($\sim0.15$ eV) but steady decrease of the optical band gaps. This drop is more prominent in the indirect bandgap estimations which also indicate the modulation of the phonon modes in the lattice via \textbf{O}-doping. It is known that dopants cause change in phonon frequency and velocity in ultra-wide-bandgap materials like AlN or diamond.\cite{wright2024acoustic, guzman2022effects} This slight decrease in bandgap from inclusion of O\textsubscript{N}\textsuperscript{+1} was also predicted by Lee \textit{et al.} through GGA theory and this is the first set of experimental data that supports such modeling.\cite{Lee2024}
\medskip

\begin{figure}[h!]
  \includegraphics[width=\linewidth]{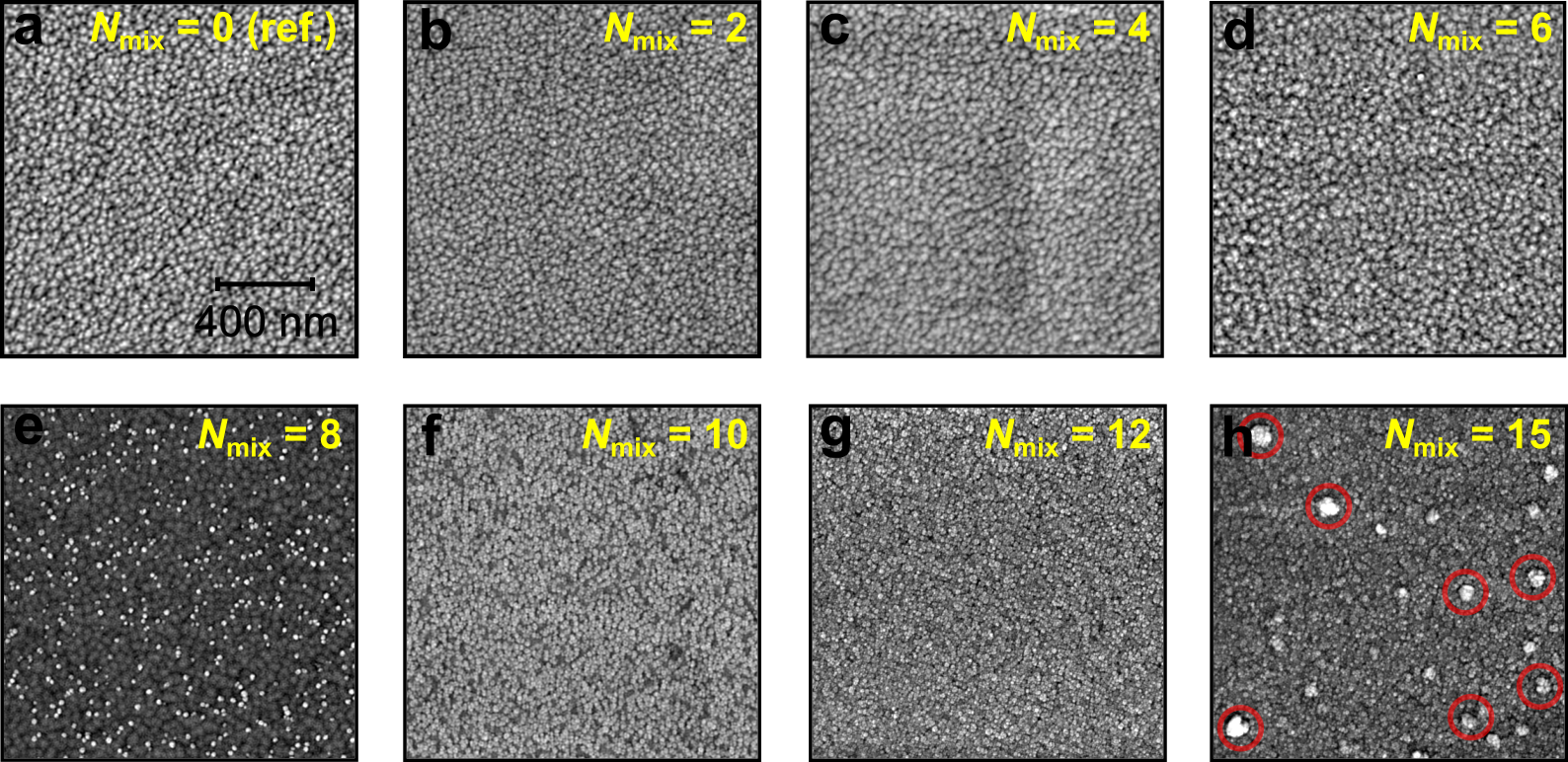}
  \caption{\small{SEM surface topography (2x2 $\mu$m\textsuperscript{2}) of the samples without (a) and with (b-h) O-doping. The red circles in (h) indicate the the presence of surface defects, e.g., oxide protrusions.}}
  \label{fig:sem01}
\end{figure}

The SEM surface microstructure observation of the \textbf{O}-doped series including the reference sample are depicted in Figure \ref{fig:sem01}. First, no significant amount of misaligned grains could be traced from the samples. The surface microstructures were identical in Figure \ref{fig:sem01}a-d, meaning up to 6 $N_\textrm{mix}$ \textbf{O}-doping, there are no evident changes in the film surface topography. However, significant changes in surface topographies were observed from 8 sccm $N_\textrm{mix}$ to 15 sccm $N_\textrm{mix}$ \textbf{O}-doped films. There are particle-like protruding grains are first apparent in the $N_\textrm{mix}$ = 8 sccm film (Figure \ref{fig:sem01}e). They are roughly 4 nm in height (AFM) and also grew in numbers as the input level of \textbf{O} reaches 10 sccm $N_\textrm{mix}$ and finally cover the complete surface almost homogeneously for the 12 sccm $N_\textrm{mix}$ specimen (Figure \ref{fig:sem01}f-g). At maximum 15 sccm $N_\textrm{mix}$ \textbf{O} doping, there are some larger agglomerations (potentially of these particles) at the film surface (some are marked in red circles at Figure \ref{fig:sem01}h). These large agglomerations are also protruding and vary in height and size (Figure \ref{fig:polarity01} k.iii). 

\begin{figure}[h!]
  \includegraphics[width=0.8\linewidth]{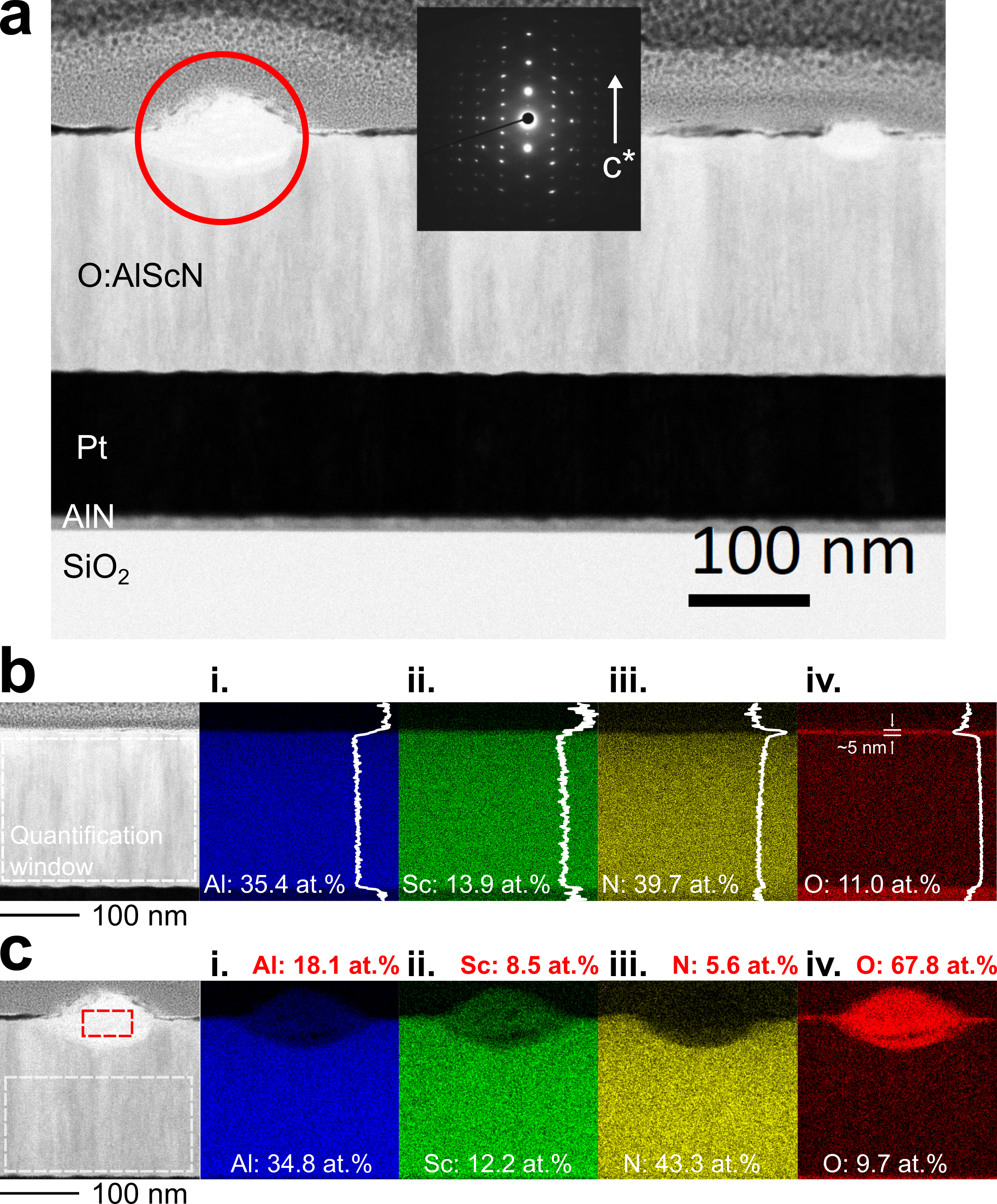}
  \centering
  \caption{\small{(a) STEM-ABF cross-section image of the \textbf{O}-doped \alscnAA\ film with $N_\textrm{mix}$ = 15 sccm. Larger and smaller surface protrusions (marked by the red circle) disseminate into the depth of the columnar film. The inset shows the SAED pattern of the bulk film - highly \textit{c}-axis fiber texture without local structural degradation. (b) STEM-EDS map of an area without surface protrusions. (c) STEM-EDS maps of an area having an oxide surface protrusion. Panels i-iv show the EDS intensity distributions of i. \textbf{Al}, ii. \textbf{Sc}, iii. \textbf{N} and iv. \textbf{O}. The averaged line profiles of the EDS intensities are overlaid on the image. The elemental composition written in red and white in panels i-iv denotes the quantified concentration of the elements inside and outside the protrusion (\%Sc $\simeq$ 27 at.\%). Note that, the EDS quantification of \textbf{N} and \textbf{O} values are affected by larger errors.}}
  \label{fig:stem01}
\end{figure}
To investigate the bulk microstructure, surface defects, and the bulk chemical composition of the highest 15 sccm $N_\textrm{mix}$ \textbf{O}-doped \alscnAA\ film, a cross-section lamella was prepared for STEM investigation. A STEM annular bright field (ABF) cross-section image at Figure \ref{fig:stem01}a displays a columnar film with two hillock-shaped protruding defects at the film surface (marked in red circles). The selected-area electron diffraction (SAED) pattern recorded from the bulk of the film confirms that the film is still highly crystalline and has \textit{c}-axis orientation as evidenced by XRD $\omega$-FWHM in Figure \ref{fig:xrd01}a (selected area diameter with the SAED aperture $\sim$ 200 nm). Chemical analysis by EDS mapping cf. Figure \ref{fig:stem01} b-c shows the elemental distribution of \textbf{Al}, \textbf{Sc}, \textbf{N}, and \textbf{O} indicating that oxygen is homogeneously distributed in the bulk of the film (Figure \ref{fig:stem01}b). In addition, an effect of post-deposition oxidation can be clearly seen in the \textbf{O} map where the native oxide thickness was measured to be around 5 nm. Further, the agglomerated surface defects were identified as some \textbf{O}-rich oxynitrides or mixtures of oxides and nitrides with reduced amount of \textbf{N}. The image also shows that these agglomerations are amorphous in nature. However, the structural origin for the defect formation remains speculative, since no significant differences in microstructure, e.g., grain orientation or phase segregation (or chemical agglomeration of \textbf{Sc}) was observed in the vicinity of the defect, which could act as weak sites for local oxidation. Below these surface defects, the bulk of the films remains unaffected, which suggests that the defects were generated randomly at a later stage in film growth. This might explain the different depth and shape of the two protrusions observed in Figure \ref{fig:stem01}a. EDS analysis also revealed no trace of Al\textsubscript{2}O\textsubscript{3} interface monolayers during growth, suggesting that the strain in the lattice comes from additional \textbf{O} rather than any seed monolayer at the bottom electrode/film interface are responsible for polarization switching in \textbf{O}-doped \alscnAA\ thin films.\cite{anggraini2020polarity}. We expect the actual \textbf{O}-content in the bulk to be lower than estimated from STEM-EDS as \textbf{O} can be introduced in the bulk during the FIB cross-section preparation and surface oxidation may occur during storage.\cite{van2008critical, schaffer2012sample} 
\pagebreak

\subsection{Effect of \textbf{O}-doping on the ferroelectric response in \textit{w-}\alscnAA}

\begin{figure}
  \centering
  \includegraphics[width=\linewidth]{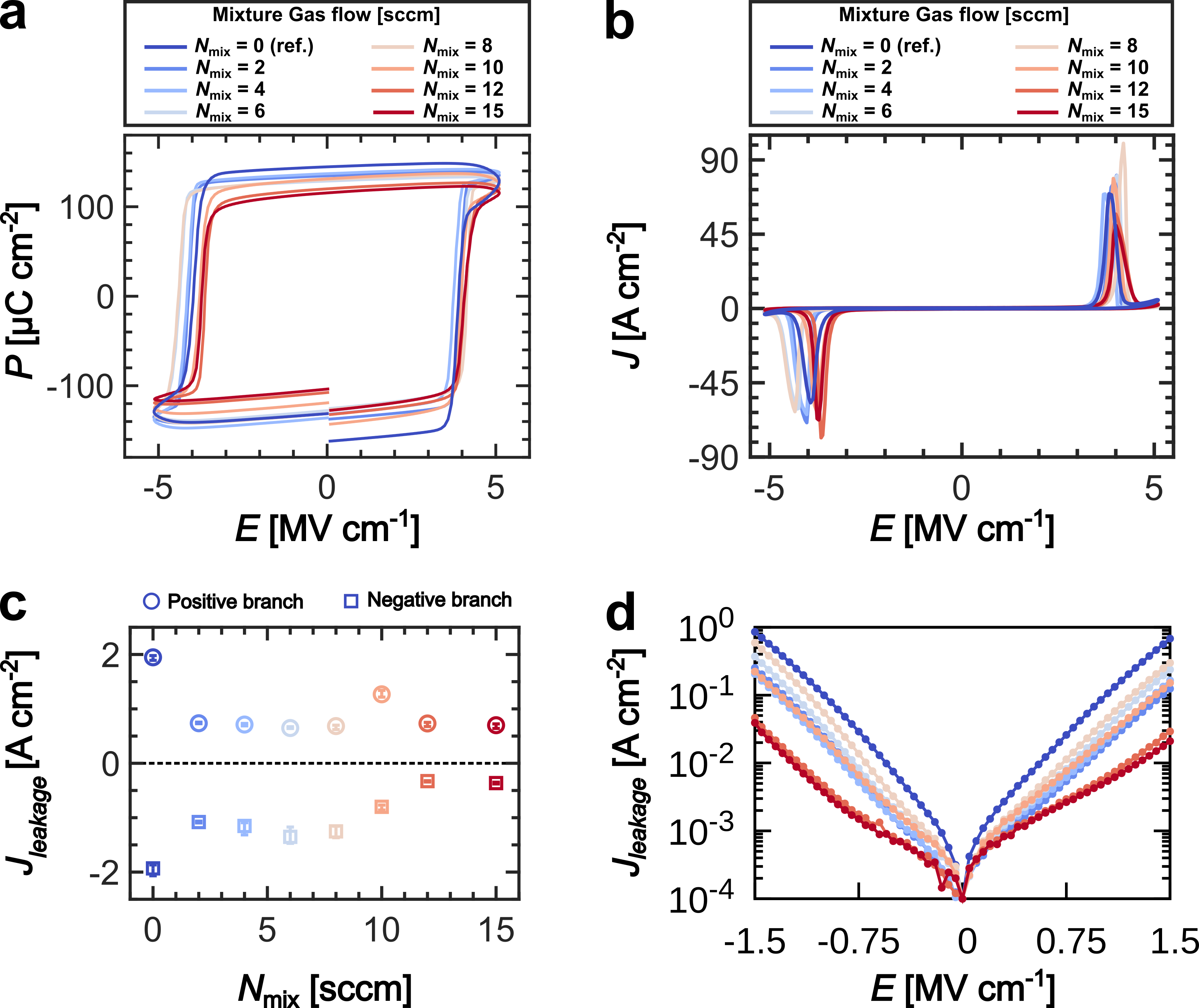}
  \caption{\small {Ferroelectric response of the Al\textsubscript{0.74}Sc\textsubscript{0.26}N without (blue, $N_\textrm{mix}$ = 0 sccm) and with O-doping ($N_\textrm{mix}$ = 2, 4, 6, 8, 10, 12, 15 sccm). (a) \textit{P-E}, (b) current density vs. electric field (\textit{J-E}) characteristics, (c) Leakage current density at 5.2 MV cm\textsuperscript{-1} (near $E_c$) at 5 kHz, (d) Steady-state leakage current density at up to 1.5 MV cm\textsuperscript{-1}. The corresponding $N_\textrm{mix}$ values for doing levels are noted in the figure legends (a) and (b)}}
  \label{fig:pe01}
\end{figure}

Figure \ref{fig:pe01}a, b shows the ferroelectric polarization and the current response of 200 nm \textit{w}-\alscnAA\ films with \textbf{O}-doping, and compares it to an undoped reference sample (blue). For comparison purpose, all the films were subjected to the same maximum field, as $E_c$ did not show any substantial trends with increasing \textbf{O}-doping and stayed in the range between 3.5 - 4.2 MV cm\textsuperscript{-1} (Figure S5, supporting information). The \textit{P-E} loops of \textbf{O}-doped samples show a sharper saturation edge (less rounded) at high electric fields for all doping concentrations which is an indicative of less leakage current contribution. Indeed, the \textit{J-E} measurements reveal a reduced leakage current tail (i.e., the current response when sweeping back from the maximum applied field) for all \textbf{O}-doped samples as shown at Figure \ref{fig:pe01}b. This observation was further verified by Figure \ref{fig:pe01}c-d where the extracted non-switching leakage current density after the coercive field (at 5 kHz) and the quasi-static leakage current measurements at lower fields. The leakage currents after $E_c$ showed a clear improvement for the \textbf{O}-doped samples compared to the undoped reference. However, this overall leakage drop can be segmented into two stages. First, at low \textbf{O}-contents, the initial four-fold drop (2x at the positive side and 2x at the negative side) which might be related to the reduction of \textbf{N}-vacancies in the bulk material by \textbf{O} atoms, due to their similar ionic radius. In addition, the formation energy for an \textbf{O} atom taking up the vacant position is lower than forming \textbf{N} vacancies. \textbf{O} incorporation in nitrides constitutes deep level impurities (DX centers).\cite{van1999defects, zhang2024dependence} It can be speculated that by replacing \textbf{N} vacancies with \textbf{O}-originating deep traps, the defect states in the band gap that can trap electrons are therefore increased, resulting in more efficient electron trapping. The second drop in leakage was prominent only at the negative branch of the \textit{P-E} loop, and occurs at the same \textbf{O} concentration at which the polarity transitions from N- to M-polar. The likely reason for this improvement is an increased barrier height at one of the interfaces due to the sign change of the bound polarization charge. Furthermore, the M-polar state of the as-grown M-polar film appears to be more resistant to the leakage current compared to the N-polar state. Similarly, the as-grown N-polar films exhibit lower leakage in the N-polar state. The drop in leakage for the as-grown M-polar films with high \textbf{O}-contents is even higher in steady-state leakage measurements at electric fields well-below the $E_c$ (up to 1.5 MV cm\textsuperscript{-1}) as depicted in Figure \ref{fig:pe01}d. At these fields, the fully M-polar films with $N_\textrm{mix}$ $\sim$ 12 and 15 sccm showed roughly an order of magnitude reduction of the leakage currents. The as-deposited loss tangent measurements showed an increase in  values with \textbf{O}-doping (Figure S6, supporting information). However, tan($\delta$) values could be expected to decrease with lower leakage current, but will also increase due to the presence of inversion domains or extrinsic defects. In our case, the as-deposited tan($\delta$) values seems to be more affected by the presence of M-polar domains and the \textbf{O}-dopants in the film.\cite{sebastian2017measurement} Therefore, our electrical measurements conclude that the presence of any systematic incorporation of oxygen atoms into the bulk via the gas phase can be used to mitigate the overall leakage current in \textit{w}-\alscnAA\ films. We speculate these improvements initiate at low \textbf{O} concentrations due to the reduction of \textbf{N} vacancies, the formation of deep level traps (DX centers) and internal polarization reversal at higher \textbf{O} concentrations within the bulk \textbf{O}-doped \textit{w}-\alscnAA.

\pagebreak

\subsection{As-deposited polarization reversal via \textbf{O}-doping}

\begin{figure}[h!]
  \includegraphics[width=0.8\linewidth]{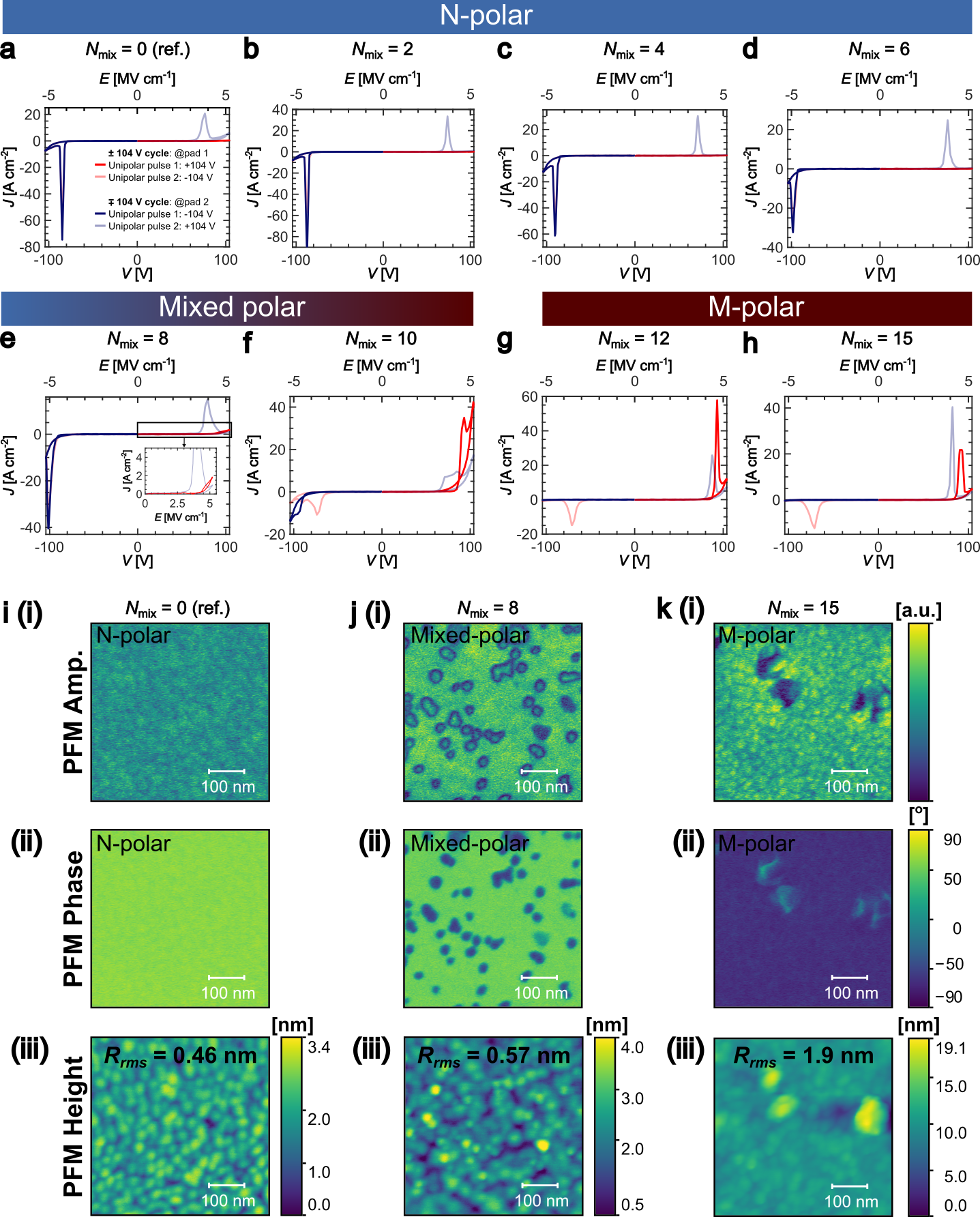}
  \caption{\small{As-deposited polarity test in 50x50 $\mu m^{2}$ pads (a) Reference \textit{w}-\alscnAA\ ($N_\textrm{mix}$ = 0 sccm), (b) $N_\textrm{mix}$ = 2 sccm, (c) $N_\textrm{mix}$ = 4 sccm, (d) $N_\textrm{mix}$ = 6 sccm, (e) $N_\textrm{mix}$ = 8 sccm, (f) $N_\textrm{mix}$ = 10 sccm, (g) $N_\textrm{mix}$ = 12 sccm, and (h) $N_\textrm{mix}$ = 15 sccm. PFM measuremtns of \textit{w}-\alscnAA\ for (i) $N_\textrm{mix}$ = 0 sccm, (j) $N_\textrm{mix}$ = 8 sccm, and (k) $N_\textrm{mix}$ = 15 sccm gas flow. The sub-figures (i), (ii) and (iii)  show the amplitude, phase and height maps correspondingly.}}
  \label{fig:polarity01}
\end{figure}

The incorporation of \textbf{O} in the III-nitrides can influence their polarity as shown for instance in AlN.\cite{akiyama2008influence} Therefore in this section, we examine the impact of \textbf{O}-doping on the as-deposited polarity of our \textit{w}-\alscnAA\ thin films. Figure \ref{fig:polarity01} a-h show the unipolar ferroelectric current-voltage (\textit{J-V}) response measurement of the samples with increasing \textbf{O}-doping concentration (the corresponding electric field values are marked at top x-axis). These measurements were conducted on two pristine capacitor structures to investigate the as-deposited polarity with voltage pulses of initial $\pm$ 104 V (shown in red @pad 1) and later $\mp$ 104 V (shown in dark-blue @pad 2). The opaque color depicts the second pulse applied to the identical electrode at the opposite field direction. During the initial $\pm$ 104 V cycle, only the negative branch exhibited a ferroelectric switching peak (panels a-d). This implies that the \textit{w-}\alscn\ at $N_\textrm{mix}$ = 0, 2, 4 and 6 sccm are fully in as-deposited N-polar state. In contrast, the \textit{J-V} loops of 8 sccm and 10 sccm $N_\textrm{mix}$ samples, as illustrated in panel e-f, exhibiting switching maxima at $\pm$ 104 V for both initial branches. Therefore, it suggests that above a certain doping threshold value achieved by adding $N_\textrm{mix}$ = 8 sccm and 10 sccm, films with as-deposited mixed M- and N-polar ferroelectric domain states are produced. Apparently, the amount of M-polar domains is even lower at $N_\textrm{mix}$ = 8 sccm compared to $N_{mix}$ = 10 sccm (lower switching current density at the +104 V cycle @pad 1, see the inset Figure \ref{fig:polarity01}f). In more highly doped $N_\textrm{mix}$ =  12 sccm and 15 sccm cases, the electrical loops displayed in Figure \ref{fig:polarity01}g-h show a switching peak on the positive branch of the $\pm$ 104 V cycle and no switching peak at the negative branch of the $\mp$ 104 V cycle. These kind of fingerprints at pad 2 are the exact opposite of a-d in pad 1, hence, indicating the presence of entirely M-polar volume at $N_\textrm{mix}$ = 12 sccm and 15 sccm. Thus, \textbf{O}-doping provides a strategy to control the as-deposited film polarity in \textit{w}-\alscnAA\ thin films by sputtering. The polarization reversal results correlate with our XRD and SEM measurements as discussed in section \ref{section1: Structural analysis with XRD, SEM, TEM} where a neck in \textbf{c} lattice parameter and changes in surface topography was observed with increased \textbf{O}-content. Furthermore, PFM measurements were performed to directly confirm the presence of polarization reversal. Subfigures i-l show the amplitude, phase and height PFM maps of the (i) reference, (j) $N_\textrm{mix}$ = 8 sccm, and (k) $N_\textrm{mix}$ = 10 sccm specimens. Comparing the sub-figure i(ii) and k (ii), the PFM phase showed a flip of almost 180\textdegree, indicating a complete N-polar state in the reference (electrical data in sub-figure a) and a complete M-polar state in the $N_\textrm{mix}$ = 15 sccm specimen (electrical data in sub-figure h). The phase map in sub-figure j(ii) shows the presence of both M- (dark blue region) and N-(green region) polar domains for the $N_\textrm{mix}$ = 8 sccm specimen (electrical data sub-figure e). The amplitude maps in sub-figures j(i) shows the presence of domain boundaries between N- and M-polar regions at $N_\textrm{mix}$ = 8 sccm. In single domain films this contrast is absent as shown in sub-figure i(i) and k(i). Furthermore, despite the formation of needle-like porous regions at the mixed polar films, the overall roughness of the reference and $N_\textrm{mix}$ = 8 sccm samples is comparable, as evidenced by the PFM height profiles at sub-figures i and j (iii). However, the roughness of the film increases significantly at $N_\textrm{mix}$ = 15 due to the formation of protruding agglomerations on the surface (sub-figure k(iii)). 
\subsection{Outline}

The incorporation of oxygen via sputtering into the bulk \textit{w-}\alscnAA\ resulted in an improvement in leakage current. Potentially, the substitution of \textbf{N}-vacancies by deep-level \textbf{O}-based traps may be the initial cause of this decrease in leakage current. In addition, the polarity of the as-deposited films was observed to be gradually inverted  from N- to M-polar when growing in oxygen-rich conditions with $N_\textrm{mix}$ $\geq$ 8 sccm. These mixed polar states at $N_\textrm{mix}$ = 8 sccm and 10 sccm were demonstrated by electrical and PFM measurements. Despite the relatively high level of \textbf{O}-incorporation inside the wurtzite-type lattice, the 0002-rocking curve FWHM of \textit{w}-\alscnAA\ showed only a $<$ 20\% increase for the highest $N_\textrm{mix}$ = 15 sccm \textbf{O}-doped film. This is unlike previously reported \textbf{O}-doping cases in \textit{w-}AlN, where a minor increase in \textbf{O}-content caused a significant loss in crystalline quality.\cite{akiyama2008influence}. The evaluation of lattice parameter changes from HRXRD also suggests intrinsic structural changes with increasing \textbf{O}-concentration and a change in film stress, which shows the tendency to relax above a threshold value (8-10 sccm) These observations might be explained by the deformation of \textbf{Sc}- and \textbf{Al}-tetrahedra by replacement of \textbf{N} atoms by \textbf{O} atoms at different doping levels which could potentially destabilize the N-polar growth mode causing a change of the as-grown polar direction. However, further insights by DFT calculations will be required in future. TEM electron diffraction and STEM-EDS investigations of our most strongly doped film also confirmed excellent [0001]-fiber texture and no chemical segregation. EDS analysis also showed no trace of Al\textsubscript{2}O\textsubscript{3} interface monolayers during growth, suggesting that the internal lattice strain rather than seed monolayer formation during material growth could be responsible for polarization switching. A homogeneous oxygen distribution within the bulk thin film is evidenced, despite the presence of larger oxide protrusions at the surface. It is clear that the oxide droplets are the result of the final stage of the \textit{w}-\alscnAA\ deposition and do not contribute to the ferroelectricity. Despite surface degradation at high oxygen concentrations, all the \textbf{O}-doped films showed lower leakage current density compared to the undoped reference samples. This reduction was roughly fourfold near the $E_c$, and reached an order of magnitude for highly doped films at low field regime. To conclude, systematic \textbf{O}-doping in sputtered \textit{w}-\alscn\ is not detrimental to the film quality or its ferroelectric property, but rather a viable alternative for improving its leakage current and regulating the as-deposited polarity.

\section{Experimental}
\subsection*{Film deposition and structuring}
The \textit{w}-\alscnAA\ films were grown on 1$\times$1 cm Pt/Si substrates by reactive pulsed DC co-sputter deposition in an Oerlikon (now Evatec) MSQ 200 multisource system. The depositions were performed at 450 \textdegree C with 7.5 sccm \textbf{Ar} and 15 sccm \textbf{N} flow. Additionally, the oxygen doping inside the film was controlled with a separate N\textsubscript{2}:O\textsubscript{2} mixture gas source connected in parallel with the pure N\textsubscript{2} source.
For electric characterization, a 100 nm Pt layer serving as top electrode was deposited and later on structured using optical lithography and ion beam etching (IBE, Oxford Instruments Ionfab 300). Note that the electrodes were deposited onto all samples at once with vacuum break, leading to surface oxidation which could not be avoided. The thickness of this oxide layer was measured $\simeq$ 5 nm (Figure \ref{fig:stem01}b.iv).\\
\subsection*{Analytical characterization}
The polarization hysteresis measurements were performed at 5 kHz (1000 averages) with a TF Analyzer 3000 (aixACCT). The leakage currents (100 averages) were measured with non-switching unipolar pulses after 1000 switching cycles. All measurements were performed by connecting the top and bottom electrode to the \textit{drive} and the \textit{sense} node of the analyzer respectively. The surface morphology and chemical composition of the doped and undoped \textit{w}-(\textbf{O})\alscnAA\ layers was examined by scanning electron microscopy on a Zeiss Gemini Ultra55 microscope (3 kV). The chemical analysis via energy dispersive X-ray spectroscopy at the SEM was performed with an acceleration voltage of 8 kV with commercial standards (Oxford Instruments). This acceleration voltage was chosen to maintain a good energy resolution for \textbf{Sc} K-$\alpha$ peaks as well as to minimize the \textbf{O}-peak contribution from X-rays generated from the buried SiO\textsubscript{2} below bottom Pt electrodes which still contributed to the 1.2 at.\% \textbf{O} for the reference. Therefore, we assume all the \textbf{O}-content reported in this article from SEM-EDS should be in the range of roughly -1 at. \% from the given value (asymmetric error bar in Figure \ref{fig:oxygencontentOO}). The HRXRD measurements were performed in a Rigaku SmartLab instrument (9 kW, Cu-k$\alpha$) equipped with a Ge(220)2x monochromator at the incident side. The in-plane measurements were conducted without the monochromator. In that case, the beam divergence for the grazing incidence were reduced with a 0.5\textdegree parallel slit collimator.  
\medskip

The \textbf{O}-content in the AlScN layers was compared using a ToF-SIMS M6 Plus instrument (IONTOF GmbH) through depth profiling in dual beam mode, employing a 2 keV Cs$^+$ sputter beam in combination with a 30 keV Bi{$_1$}$^+$ primary beam for crater analysis. Due to the absence of standards, only relative comparisons were conducted.

UV-Vis measurements were performed with a Cary5000 UV-Vis-NIR Spectrophotometer probing from 800-200 nm wavelengths with a step size of one nanometer. Prior to taking data, the system was normalized with baseline (through beam) and zero (blocked beam) measurements. All transmission data of samples were then normalized to a bare substrate measurement to correct for any UV absorption of the substrate such that
\begin{align}
T(\lambda) = T_{sample}(\lambda) / T_{substrate}(\lambda)
\end{align}
We followed the standard analysis practice of as outlined below\cite{Jaramillo_Tauc, Vachhani}. Analysis for the best estimation of the direct and indirect bandgaps was used with the Tauc relation,
\begin{align}
   \alpha hv \propto (hv-Eg)^x 
\end{align}
Where x = ½ or 2 for direct or indirect bandgap respectively. $\alpha$ is the absorption coefficient calculated by,
\begin{align}
    \alpha = -ln(10)log(T)/d
\end{align}
Plotting \(\alpha hv^{1/x}\) vs \(hv\) yields two linear regimes. By extrapolating these linear trends to their intercept, the band gaps were estimated for \alscnAA\ with different \textbf{O}-doping amounts. 
However, these measurements only provide estimations of the optical bandgap which might differ slightly from the electronic band gap of the material. More information on this process and its limitations are summarized by Klein et al.\cite{Klein2023Tauc}

\medskip
STEM analysis and STEM-EDS measurements were performed using a JEOL NEOARM microscope at 200 kV equipped with a dual windowless EDS detector system. The PFM measurements were performed in a Park Systems NX20, off resonance PFM at 17 kHz (500x500 nm, with an adaptive rate of 0.05 to 0.5 Hz). For the measurement, the sample was grounded and The tip was drive from $\pm$10 V. 

%%%%%%%%%%%%%%%%%%%%%%%%%%%%%%%%%%%%%%%%%%%%%%%%%%%%%%%%%%%%%%%%%%%%%
%% The "Acknowledgement" section can be given in all manuscript
%% classes.  This should be given within the "acknowledgement"
%% environment, which will make the correct section or running title.
%%%%%%%%%%%%%%%%%%%%%%%%%%%%%%%%%%%%%%%%%%%%%%%%%%%%%%%%%%%%%%%%%%%%%
\begin{acknowledgement}

The authors thank Christin Szillus for \textsf{TEM} sample preparation and Dr. Fabio Aldo Kraft for the help in \textsf{UV-Vis} measurements.

\end{acknowledgement}

%%%%%%%%%%%%%%%%%%%%%%%%%%%%%%%%%%%%%%%%%%%%%%%%%%%%%%%%%%%%%%%%%%%%%
%% The same is true for Supporting Information, which should use the
%% suppinfo environment.
%%%%%%%%%%%%%%%%%%%%%%%%%%%%%%%%%%%%%%%%%%%%%%%%%%%%%%%%%%%%%%%%%%%%%
\begin{suppinfo}

The Supporting Information is available free of charge at

\end{suppinfo}

%%%%%%%%%%%%%%%%%%%%%%%%%%%%%%%%%%%%%%%%%%%%%%%%%%%%%%%%%%%%%%%%%%%%%
%% The appropriate \bibliography command should be placed here.
%% Notice that the class file automatically sets \bibliographystyle
%% and also names the section correctly.
%%%%%%%%%%%%%%%%%%%%%%%%%%%%%%%%%%%%%%%%%%%%%%%%%%%%%%%%%%%%%%%%%%%%%
\bibliography{achemso-demo}

\providecommand{\latin}[1]{#1}
\makeatletter
\providecommand{\doi}
  {\begingroup\let\do\@makeother\dospecials
  \catcode`\{=1 \catcode`\}=2 \doi@aux}
\providecommand{\doi@aux}[1]{\endgroup\texttt{#1}}
\makeatother
\providecommand*\mcitethebibliography{\thebibliography}
\csname @ifundefined\endcsname{endmcitethebibliography}  {\let\endmcitethebibliography\endthebibliography}{}
\begin{mcitethebibliography}{48}
\providecommand*\natexlab[1]{#1}
\providecommand*\mciteSetBstSublistMode[1]{}
\providecommand*\mciteSetBstMaxWidthForm[2]{}
\providecommand*\mciteBstWouldAddEndPuncttrue
  {\def\EndOfBibitem{\unskip.}}
\providecommand*\mciteBstWouldAddEndPunctfalse
  {\let\EndOfBibitem\relax}
\providecommand*\mciteSetBstMidEndSepPunct[3]{}
\providecommand*\mciteSetBstSublistLabelBeginEnd[3]{}
\providecommand*\EndOfBibitem{}
\mciteSetBstSublistMode{f}
\mciteSetBstMaxWidthForm{subitem}{(\alph{mcitesubitemcount})}
\mciteSetBstSublistLabelBeginEnd
  {\mcitemaxwidthsubitemform\space}
  {\relax}
  {\relax}

\bibitem[Fichtner \latin{et~al.}(2019)Fichtner, Wolff, Lofink, Kienle, and Wagner]{fichtner2019alscn}
Fichtner,~S.; Wolff,~N.; Lofink,~F.; Kienle,~L.; Wagner,~B. AlScN: A III-V semiconductor based ferroelectric. \emph{Journal of Applied Physics} \textbf{2019}, \emph{125}\relax
\mciteBstWouldAddEndPuncttrue
\mciteSetBstMidEndSepPunct{\mcitedefaultmidpunct}
{\mcitedefaultendpunct}{\mcitedefaultseppunct}\relax
\EndOfBibitem
\bibitem[Leone \latin{et~al.}(2020)Leone, Ligl, Manz, Kirste, Fuchs, Menner, Prescher, Wiegert, {\v{Z}}ukauskait{\.e}, Quay, \latin{et~al.} others]{leone2020metal}
Leone,~S.; Ligl,~J.; Manz,~C.; Kirste,~L.; Fuchs,~T.; Menner,~H.; Prescher,~M.; Wiegert,~J.; {\v{Z}}ukauskait{\.e},~A.; Quay,~R.; others Metal-organic chemical vapor deposition of aluminum scandium nitride. \emph{physica status solidi (RRL)--Rapid Research Letters} \textbf{2020}, \emph{14}, 1900535\relax
\mciteBstWouldAddEndPuncttrue
\mciteSetBstMidEndSepPunct{\mcitedefaultmidpunct}
{\mcitedefaultendpunct}{\mcitedefaultseppunct}\relax
\EndOfBibitem
\bibitem[Wang \latin{et~al.}(2021)Wang, Wang, Vu, Chiang, Heron, and Mi]{wang2021fully}
Wang,~P.; Wang,~D.; Vu,~N.~M.; Chiang,~T.; Heron,~J.~T.; Mi,~Z. Fully epitaxial ferroelectric ScAlN grown by molecular beam epitaxy. \emph{Applied Physics Letters} \textbf{2021}, \emph{118}\relax
\mciteBstWouldAddEndPuncttrue
\mciteSetBstMidEndSepPunct{\mcitedefaultmidpunct}
{\mcitedefaultendpunct}{\mcitedefaultseppunct}\relax
\EndOfBibitem
\bibitem[Kim \latin{et~al.}(2023)Kim, Oh, Fiagbenu, Zheng, Musavigharavi, Kumar, Trainor, Aljarb, Wan, Kim, \latin{et~al.} others]{kim2023scalable}
Kim,~K.-H.; Oh,~S.; Fiagbenu,~M. M.~A.; Zheng,~J.; Musavigharavi,~P.; Kumar,~P.; Trainor,~N.; Aljarb,~A.; Wan,~Y.; Kim,~H.~M.; others Scalable CMOS back-end-of-line-compatible AlScN/two-dimensional channel ferroelectric field-effect transistors. \emph{Nature nanotechnology} \textbf{2023}, \emph{18}, 1044--1050\relax
\mciteBstWouldAddEndPuncttrue
\mciteSetBstMidEndSepPunct{\mcitedefaultmidpunct}
{\mcitedefaultendpunct}{\mcitedefaultseppunct}\relax
\EndOfBibitem
\bibitem[Mikolajick \latin{et~al.}(2021)Mikolajick, Slesazeck, Mulaosmanovic, Park, Fichtner, Lomenzo, Hoffmann, and Schroeder]{mikolajick2021next}
Mikolajick,~T.; Slesazeck,~S.; Mulaosmanovic,~H.; Park,~M.; Fichtner,~S.; Lomenzo,~P.; Hoffmann,~M.; Schroeder,~U. Next generation ferroelectric materials for semiconductor process integration and their applications. \emph{Journal of Applied Physics} \textbf{2021}, \emph{129}\relax
\mciteBstWouldAddEndPuncttrue
\mciteSetBstMidEndSepPunct{\mcitedefaultmidpunct}
{\mcitedefaultendpunct}{\mcitedefaultseppunct}\relax
\EndOfBibitem
\bibitem[Gremmel and Fichtner(2024)Gremmel, and Fichtner]{gremmel2024interplay}
Gremmel,~M.; Fichtner,~S. The interplay between imprint, wake-up, and domains in ferroelectric Al0. 70Sc0. 30N. \emph{Journal of Applied Physics} \textbf{2024}, \emph{135}\relax
\mciteBstWouldAddEndPuncttrue
\mciteSetBstMidEndSepPunct{\mcitedefaultmidpunct}
{\mcitedefaultendpunct}{\mcitedefaultseppunct}\relax
\EndOfBibitem
\bibitem[Islam \latin{et~al.}(2021)Islam, Wolff, Yassine, Sch\"{o}nweger, Christian, Kohlstedt, Ambacher, Lofink, Kienle, and Fichtner]{islam2021exceptional}
Islam,~M.~R.; Wolff,~N.; Yassine,~M.; Sch\"{o}nweger,~G.; Christian,~B.; Kohlstedt,~H.; Ambacher,~O.; Lofink,~F.; Kienle,~L.; Fichtner,~S. On the exceptional temperature stability of ferroelectric Al1-xScxN thin films. \emph{Applied Physics Letters} \textbf{2021}, \emph{118}\relax
\mciteBstWouldAddEndPuncttrue
\mciteSetBstMidEndSepPunct{\mcitedefaultmidpunct}
{\mcitedefaultendpunct}{\mcitedefaultseppunct}\relax
\EndOfBibitem
\bibitem[Wang \latin{et~al.}(2023)Wang, Wang, Mondal, Hu, Wu, Ma, and Mi]{wang2023ferroelectric}
Wang,~P.; Wang,~D.; Mondal,~S.; Hu,~M.; Wu,~Y.; Ma,~T.; Mi,~Z. Ferroelectric nitride heterostructures on CMOS compatible molybdenum for synaptic memristors. \emph{ACS Applied Materials \& Interfaces} \textbf{2023}, \emph{15}, 18022--18031\relax
\mciteBstWouldAddEndPuncttrue
\mciteSetBstMidEndSepPunct{\mcitedefaultmidpunct}
{\mcitedefaultendpunct}{\mcitedefaultseppunct}\relax
\EndOfBibitem
\bibitem[Drury \latin{et~al.}(2022)Drury, Yazawa, Zakutayev, Hanrahan, and Brennecka]{drury2022high}
Drury,~D.; Yazawa,~K.; Zakutayev,~A.; Hanrahan,~B.; Brennecka,~G. High-temperature ferroelectric behavior of Al0. 7Sc0. 3N. \emph{Micromachines} \textbf{2022}, \emph{13}, 887\relax
\mciteBstWouldAddEndPuncttrue
\mciteSetBstMidEndSepPunct{\mcitedefaultmidpunct}
{\mcitedefaultendpunct}{\mcitedefaultseppunct}\relax
\EndOfBibitem
\bibitem[Drury(2023)]{alma998135459402341}
Drury,~I.,~Daniel~E. Toward high operating temperature AlN-based ferroelectric random access memory. 2023\relax
\mciteBstWouldAddEndPuncttrue
\mciteSetBstMidEndSepPunct{\mcitedefaultmidpunct}
{\mcitedefaultendpunct}{\mcitedefaultseppunct}\relax
\EndOfBibitem
\bibitem[Streicher \latin{et~al.}(2023)Streicher, Leone, Kirste, Manz, Stra{\v{n}}{\'a}k, Prescher, Waltereit, Mikulla, Quay, and Ambacher]{streicher2023enhanced}
Streicher,~I.; Leone,~S.; Kirste,~L.; Manz,~C.; Stra{\v{n}}{\'a}k,~P.; Prescher,~M.; Waltereit,~P.; Mikulla,~M.; Quay,~R.; Ambacher,~O. Enhanced AlScN/GaN heterostructures grown with a novel precursor by metal--organic chemical vapor deposition. \emph{physica status solidi (RRL)--Rapid Research Letters} \textbf{2023}, \emph{17}, 2200387\relax
\mciteBstWouldAddEndPuncttrue
\mciteSetBstMidEndSepPunct{\mcitedefaultmidpunct}
{\mcitedefaultendpunct}{\mcitedefaultseppunct}\relax
\EndOfBibitem
\bibitem[Krause \latin{et~al.}(2022)Krause, Streicher, Waltereit, Kirste, Br{\"u}ckner, and Leone]{krause2022alscn}
Krause,~S.; Streicher,~I.; Waltereit,~P.; Kirste,~L.; Br{\"u}ckner,~P.; Leone,~S. AlScN/GaN HEMTs grown by metal-organic chemical vapor deposition with 8.4 W/mm output power and 48\% power-added efficiency at 30 GHz. \emph{IEEE Electron Device Letters} \textbf{2022}, \emph{44}, 17--20\relax
\mciteBstWouldAddEndPuncttrue
\mciteSetBstMidEndSepPunct{\mcitedefaultmidpunct}
{\mcitedefaultendpunct}{\mcitedefaultseppunct}\relax
\EndOfBibitem
\bibitem[Hayden \latin{et~al.}(2021)Hayden, Hossain, Xiong, Ferri, Zhu, Imperatore, Giebink, Trolier-McKinstry, Dabo, and Maria]{hayden2021ferroelectricity}
Hayden,~J.; Hossain,~M.~D.; Xiong,~Y.; Ferri,~K.; Zhu,~W.; Imperatore,~M.~V.; Giebink,~N.; Trolier-McKinstry,~S.; Dabo,~I.; Maria,~J.-P. Ferroelectricity in boron-substituted aluminum nitride thin films. \emph{Physical Review Materials} \textbf{2021}, \emph{5}, 044412\relax
\mciteBstWouldAddEndPuncttrue
\mciteSetBstMidEndSepPunct{\mcitedefaultmidpunct}
{\mcitedefaultendpunct}{\mcitedefaultseppunct}\relax
\EndOfBibitem
\bibitem[Wang \latin{et~al.}(2021)Wang, Wang, Wang, and Mi]{wang2021fullygascn}
Wang,~D.; Wang,~P.; Wang,~B.; Mi,~Z. Fully epitaxial ferroelectric ScGaN grown on GaN by molecular beam epitaxy. \emph{Applied Physics Letters} \textbf{2021}, \emph{119}\relax
\mciteBstWouldAddEndPuncttrue
\mciteSetBstMidEndSepPunct{\mcitedefaultmidpunct}
{\mcitedefaultendpunct}{\mcitedefaultseppunct}\relax
\EndOfBibitem
\bibitem[Wang \latin{et~al.}(2023)Wang, Mondal, Liu, Hu, Wang, Yang, Wang, Xiao, Wu, Ma, \latin{et~al.} others]{wang2023yaln}
Wang,~D.; Mondal,~S.; Liu,~J.; Hu,~M.; Wang,~P.; Yang,~S.; Wang,~D.; Xiao,~Y.; Wu,~Y.; Ma,~T.; others Ferroelectric YAlN grown by molecular beam epitaxy. \emph{Applied Physics Letters} \textbf{2023}, \emph{123}\relax
\mciteBstWouldAddEndPuncttrue
\mciteSetBstMidEndSepPunct{\mcitedefaultmidpunct}
{\mcitedefaultendpunct}{\mcitedefaultseppunct}\relax
\EndOfBibitem
\bibitem[Islam \latin{et~al.}(2023)Islam, Scho\"{o}nweger, Wolff, Petraru, Kohlstedt, Fichtner, and Kienle]{islam2023comparative}
Islam,~M.~R.; Scho\"{o}nweger,~G.; Wolff,~N.; Petraru,~A.; Kohlstedt,~H.; Fichtner,~S.; Kienle,~L. A Comparative Study of Pt/Al0. 72Sc0. 28N/Pt-Based Thin-Film Metal-Ferroelectric-Metal Capacitors on GaN and Si Substrates. \emph{ACS Applied Materials \& Interfaces} \textbf{2023}, \emph{15}, 41606--41613\relax
\mciteBstWouldAddEndPuncttrue
\mciteSetBstMidEndSepPunct{\mcitedefaultmidpunct}
{\mcitedefaultendpunct}{\mcitedefaultseppunct}\relax
\EndOfBibitem
\bibitem[Sch\"{o}nweger \latin{et~al.}(2022)Sch\"{o}nweger, Petraru, Islam, Wolff, Haas, Hammud, Koch, Kienle, Kohlstedt, and Fichtner]{schonweger2022fully}
Sch\"{o}nweger,~G.; Petraru,~A.; Islam,~M.~R.; Wolff,~N.; Haas,~B.; Hammud,~A.; Koch,~C.; Kienle,~L.; Kohlstedt,~H.; Fichtner,~S. From fully strained to relaxed: epitaxial ferroelectric Al1-xScxN for III-N technology. \emph{Advanced Functional Materials} \textbf{2022}, \emph{32}, 2109632\relax
\mciteBstWouldAddEndPuncttrue
\mciteSetBstMidEndSepPunct{\mcitedefaultmidpunct}
{\mcitedefaultendpunct}{\mcitedefaultseppunct}\relax
\EndOfBibitem
\bibitem[Yang \latin{et~al.}(2024)Yang, Tang, Sun, Chen, Jiang, Zhang, and Dong]{yang2024emerging}
Yang,~D.-P.; Tang,~X.-G.; Sun,~Q.-J.; Chen,~J.-Y.; Jiang,~Y.-P.; Zhang,~D.; Dong,~H.-F. Emerging ferroelectric materials ScAlN: applications and prospects in memristors. \emph{Materials Horizons} \textbf{2024}, \relax
\mciteBstWouldAddEndPunctfalse
\mciteSetBstMidEndSepPunct{\mcitedefaultmidpunct}
{}{\mcitedefaultseppunct}\relax
\EndOfBibitem
\bibitem[Tsai \latin{et~al.}(2021)Tsai, Hoshii, Wakabayashi, Tsutsui, Chung, Chang, and Kakushima]{tsai2021room}
Tsai,~S.-L.; Hoshii,~T.; Wakabayashi,~H.; Tsutsui,~K.; Chung,~T.-K.; Chang,~E.~Y.; Kakushima,~K. Room-temperature deposition of a poling-free ferroelectric AlScN film by reactive sputtering. \emph{Applied Physics Letters} \textbf{2021}, \emph{118}\relax
\mciteBstWouldAddEndPuncttrue
\mciteSetBstMidEndSepPunct{\mcitedefaultmidpunct}
{\mcitedefaultendpunct}{\mcitedefaultseppunct}\relax
\EndOfBibitem
\bibitem[Wolff \latin{et~al.}(2023)Wolff, Sch{\"o}nweger, Streicher, Islam, Braun, Stra{\v{n}}{\'a}k, Kirste, Prescher, Lotnyk, Kohlstedt, \latin{et~al.} others]{wolff2023demonstration}
Wolff,~N.; Sch{\"o}nweger,~G.; Streicher,~I.; Islam,~M.~R.; Braun,~N.; Stra{\v{n}}{\'a}k,~P.; Kirste,~L.; Prescher,~M.; Lotnyk,~A.; Kohlstedt,~H.; others Demonstration and STEM Analysis of Ferroelectric Switching in MOCVD-Grown Single Crystalline Al0. 85Sc0. 15N. \emph{Advanced Physics Research} \textbf{2023}, 2300113\relax
\mciteBstWouldAddEndPuncttrue
\mciteSetBstMidEndSepPunct{\mcitedefaultmidpunct}
{\mcitedefaultendpunct}{\mcitedefaultseppunct}\relax
\EndOfBibitem
\bibitem[Wang \latin{et~al.}(2020)Wang, Laleyan, Pandey, Sun, and Mi]{wang2020molecular}
Wang,~P.; Laleyan,~D.~A.; Pandey,~A.; Sun,~Y.; Mi,~Z. Molecular beam epitaxy and characterization of wurtzite ScxAl1- xN. \emph{Applied Physics Letters} \textbf{2020}, \emph{116}\relax
\mciteBstWouldAddEndPuncttrue
\mciteSetBstMidEndSepPunct{\mcitedefaultmidpunct}
{\mcitedefaultendpunct}{\mcitedefaultseppunct}\relax
\EndOfBibitem
\bibitem[Sch{\"o}nweger \latin{et~al.}(2023)Sch{\"o}nweger, Wolff, Islam, Gremmel, Petraru, Kienle, Kohlstedt, and Fichtner]{schonweger2023grain}
Sch{\"o}nweger,~G.; Wolff,~N.; Islam,~M.~R.; Gremmel,~M.; Petraru,~A.; Kienle,~L.; Kohlstedt,~H.; Fichtner,~S. In-grain ferroelectric switching in sub-5 nm thin Al0. 74Sc0. 26N films at 1 V. \emph{Advanced Science} \textbf{2023}, \emph{10}, 2302296\relax
\mciteBstWouldAddEndPuncttrue
\mciteSetBstMidEndSepPunct{\mcitedefaultmidpunct}
{\mcitedefaultendpunct}{\mcitedefaultseppunct}\relax
\EndOfBibitem
\bibitem[Akiyama \latin{et~al.}(2009)Akiyama, Kamohara, Kano, Teshigahara, Takeuchi, and Kawahara]{akiyama2009enhancement}
Akiyama,~M.; Kamohara,~T.; Kano,~K.; Teshigahara,~A.; Takeuchi,~Y.; Kawahara,~N. Enhancement of piezoelectric response in scandium aluminum nitride alloy thin films prepared by dual reactive cosputtering. \emph{Advanced Materials} \textbf{2009}, \emph{21}, 593--596\relax
\mciteBstWouldAddEndPuncttrue
\mciteSetBstMidEndSepPunct{\mcitedefaultmidpunct}
{\mcitedefaultendpunct}{\mcitedefaultseppunct}\relax
\EndOfBibitem
\bibitem[Akiyama \latin{et~al.}(2009)Akiyama, Kano, and Teshigahara]{akiyama2009influence}
Akiyama,~M.; Kano,~K.; Teshigahara,~A. Influence of growth temperature and scandium concentration on piezoelectric response of scandium aluminum nitride alloy thin films. \emph{Applied Physics Letters} \textbf{2009}, \emph{95}\relax
\mciteBstWouldAddEndPuncttrue
\mciteSetBstMidEndSepPunct{\mcitedefaultmidpunct}
{\mcitedefaultendpunct}{\mcitedefaultseppunct}\relax
\EndOfBibitem
\bibitem[Sch{\"o}nweger \latin{et~al.}(2023)Sch{\"o}nweger, Islam, and Fichtner]{schonweger2023structural}
Sch{\"o}nweger,~G.; Islam,~M.~R.; Fichtner,~S. Structural and ferroelectric properties of Al1- xScxN. \textbf{2023}, \relax
\mciteBstWouldAddEndPunctfalse
\mciteSetBstMidEndSepPunct{\mcitedefaultmidpunct}
{}{\mcitedefaultseppunct}\relax
\EndOfBibitem
\bibitem[Geng \latin{et~al.}(2021)Geng, Chen, Pan, Qiao, He, Mu, Hou, and Chou]{geng2021improved}
Geng,~W.; Chen,~X.; Pan,~L.; Qiao,~X.; He,~J.; Mu,~J.; Hou,~X.; Chou,~X. Improved crystallization, domain, and ferroelectricity by controlling lead/oxygen vacancies in Mn-doped PZT thin films. \emph{Materials Characterization} \textbf{2021}, \emph{176}, 111131\relax
\mciteBstWouldAddEndPuncttrue
\mciteSetBstMidEndSepPunct{\mcitedefaultmidpunct}
{\mcitedefaultendpunct}{\mcitedefaultseppunct}\relax
\EndOfBibitem
\bibitem[Zhang \latin{et~al.}(2024)Zhang, Yang, Liang, Liu, Hou, Liu, Zheng, Liu, and Zhao]{zhang2024dependence}
Zhang,~Y.; Yang,~J.; Liang,~F.; Liu,~Z.; Hou,~Y.; Liu,~B.; Zheng,~F.; Liu,~X.; Zhao,~D. Dependence of oxygen impurity concentration in AlN on the surface roughness during growth. \emph{Journal of Applied Physics} \textbf{2024}, \emph{135}\relax
\mciteBstWouldAddEndPuncttrue
\mciteSetBstMidEndSepPunct{\mcitedefaultmidpunct}
{\mcitedefaultendpunct}{\mcitedefaultseppunct}\relax
\EndOfBibitem
\bibitem[Chen \latin{et~al.}(2024)Chen, Nishida, Tsai, Hoshii, Tsutsui, Wakabayashi, Chang, and Kakushima]{chen2024oxygen}
Chen,~S.-M.; Nishida,~H.; Tsai,~S.-L.; Hoshii,~T.; Tsutsui,~K.; Wakabayashi,~H.; Chang,~E.~Y.; Kakushima,~K. Oxygen-atom Incorporated Ferroelectric AIScN Capacitors for Multi-level Operation. \emph{IEEE Electron Device Letters} \textbf{2024}, \relax
\mciteBstWouldAddEndPunctfalse
\mciteSetBstMidEndSepPunct{\mcitedefaultmidpunct}
{}{\mcitedefaultseppunct}\relax
\EndOfBibitem
\bibitem[Zhu \latin{et~al.}(2020)Zhu, Liu, Lu, Xu, Qi, and Zhang]{zhu2020effect}
Zhu,~X.; Liu,~K.; Lu,~Z.; Xu,~Y.; Qi,~S.; Zhang,~G. Effect of oxygen atoms on graphene: Adsorption and doping. \emph{Physica E: Low-dimensional Systems and Nanostructures} \textbf{2020}, \emph{117}, 113827\relax
\mciteBstWouldAddEndPuncttrue
\mciteSetBstMidEndSepPunct{\mcitedefaultmidpunct}
{\mcitedefaultendpunct}{\mcitedefaultseppunct}\relax
\EndOfBibitem
\bibitem[Slater(1964)]{slater1964atomic}
Slater,~J.~C. Atomic radii in crystals. \emph{The Journal of Chemical Physics} \textbf{1964}, \emph{41}, 3199--3204\relax
\mciteBstWouldAddEndPuncttrue
\mciteSetBstMidEndSepPunct{\mcitedefaultmidpunct}
{\mcitedefaultendpunct}{\mcitedefaultseppunct}\relax
\EndOfBibitem
\bibitem[Lee \latin{et~al.}(2024)Lee, Din, Brennecka, and Gorai]{Lee2024}
Lee,~C.-W.; Din,~N.~U.; Brennecka,~G.~L.; Gorai,~P. Defects and oxygen impurities in ferroelectric wurtzite Al1-xScxN alloys. \emph{Applied Physics Letters} \textbf{2024}, \emph{125}, 022901\relax
\mciteBstWouldAddEndPuncttrue
\mciteSetBstMidEndSepPunct{\mcitedefaultmidpunct}
{\mcitedefaultendpunct}{\mcitedefaultseppunct}\relax
\EndOfBibitem
\bibitem[Gordon \latin{et~al.}(2014)Gordon, Lyons, Janotti, and Van~de Walle]{gordon2014hybrid}
Gordon,~L.; Lyons,~J.; Janotti,~A.; Van~de Walle,~C. Hybrid functional calculations of DX centers in AlN and GaN. \emph{Physical Review B} \textbf{2014}, \emph{89}, 085204\relax
\mciteBstWouldAddEndPuncttrue
\mciteSetBstMidEndSepPunct{\mcitedefaultmidpunct}
{\mcitedefaultendpunct}{\mcitedefaultseppunct}\relax
\EndOfBibitem
\bibitem[Yang \latin{et~al.}(2014)Yang, Chang, Hsiao, Lee, and Lou]{yang2014influence}
Yang,~Y.-C.; Chang,~C.-T.; Hsiao,~Y.-C.; Lee,~J.-W.; Lou,~B.-S. Influence of high power impulse magnetron sputtering pulse parameters on the properties of aluminum nitride coatings. \emph{Surface and Coatings Technology} \textbf{2014}, \emph{259}, 219--231\relax
\mciteBstWouldAddEndPuncttrue
\mciteSetBstMidEndSepPunct{\mcitedefaultmidpunct}
{\mcitedefaultendpunct}{\mcitedefaultseppunct}\relax
\EndOfBibitem
\bibitem[Chen \latin{et~al.}(2024)Chen, Hoshii, Wakabayashi, Tsutsui, Chang, and Kakushima]{chen2024reactive}
Chen,~S.-M.; Hoshii,~T.; Wakabayashi,~H.; Tsutsui,~K.; Chang,~E.~Y.; Kakushima,~K. Reactive sputtering of ferroelectric AlScN films with H2 gas flow for endurance improvement. \emph{Japanese Journal of Applied Physics} \textbf{2024}, \emph{63}, 03SP45\relax
\mciteBstWouldAddEndPuncttrue
\mciteSetBstMidEndSepPunct{\mcitedefaultmidpunct}
{\mcitedefaultendpunct}{\mcitedefaultseppunct}\relax
\EndOfBibitem
\bibitem[Moram \latin{et~al.}(2008)Moram, Barber, and Humphreys]{moram2008effect}
Moram,~M.; Barber,~Z.; Humphreys,~C. The effect of oxygen incorporation in sputtered scandium nitride films. \emph{Thin Solid Films} \textbf{2008}, \emph{516}, 8569--8572\relax
\mciteBstWouldAddEndPuncttrue
\mciteSetBstMidEndSepPunct{\mcitedefaultmidpunct}
{\mcitedefaultendpunct}{\mcitedefaultseppunct}\relax
\EndOfBibitem
\bibitem[Luo(2007)]{luo2007}
Luo,~Y.-R. \emph{Comprehensive {Handbook} of {Chemical} {Bond} {Energies}}; CRC Press: Boca Raton, 2007\relax
\mciteBstWouldAddEndPuncttrue
\mciteSetBstMidEndSepPunct{\mcitedefaultmidpunct}
{\mcitedefaultendpunct}{\mcitedefaultseppunct}\relax
\EndOfBibitem
\bibitem[Wright \latin{et~al.}(2024)Wright, Mudiyanselage, Guzman, Fu, Teeter, Da, Kargar, Fu, and Balandin]{wright2024acoustic}
Wright,~D.; Mudiyanselage,~D.~H.; Guzman,~E.; Fu,~X.; Teeter,~J.; Da,~B.; Kargar,~F.; Fu,~H.; Balandin,~A.~A. Acoustic and optical phonon frequencies and acoustic phonon velocities in Si-doped AlN thin films. \emph{Applied Physics Letters} \textbf{2024}, \emph{125}\relax
\mciteBstWouldAddEndPuncttrue
\mciteSetBstMidEndSepPunct{\mcitedefaultmidpunct}
{\mcitedefaultendpunct}{\mcitedefaultseppunct}\relax
\EndOfBibitem
\bibitem[Guzman \latin{et~al.}(2022)Guzman, Kargar, Angeles, Meidanshahi, Grotjohn, Hardy, Muehle, Wilson, Goodnick, and Balandin]{guzman2022effects}
Guzman,~E.; Kargar,~F.; Angeles,~F.; Meidanshahi,~R.~V.; Grotjohn,~T.; Hardy,~A.; Muehle,~M.; Wilson,~R.~B.; Goodnick,~S.~M.; Balandin,~A.~A. Effects of boron doping on the bulk and surface acoustic phonons in single-crystal diamond. \emph{ACS Applied Materials \& Interfaces} \textbf{2022}, \emph{14}, 42223--42231\relax
\mciteBstWouldAddEndPuncttrue
\mciteSetBstMidEndSepPunct{\mcitedefaultmidpunct}
{\mcitedefaultendpunct}{\mcitedefaultseppunct}\relax
\EndOfBibitem
\bibitem[Anggraini \latin{et~al.}(2020)Anggraini, Uehara, Hirata, Yamada, and Akiyama]{anggraini2020polarity}
Anggraini,~S.~A.; Uehara,~M.; Hirata,~K.; Yamada,~H.; Akiyama,~M. Polarity inversion of aluminum nitride thin films by using Si and MgSi dopants. \emph{Scientific Reports} \textbf{2020}, \emph{10}, 4369\relax
\mciteBstWouldAddEndPuncttrue
\mciteSetBstMidEndSepPunct{\mcitedefaultmidpunct}
{\mcitedefaultendpunct}{\mcitedefaultseppunct}\relax
\EndOfBibitem
\bibitem[Van~Dorp and Hagen(2008)Van~Dorp, and Hagen]{van2008critical}
Van~Dorp,~W.; Hagen,~C.~W. A critical literature review of focused electron beam induced deposition. \emph{Journal of Applied Physics} \textbf{2008}, \emph{104}\relax
\mciteBstWouldAddEndPuncttrue
\mciteSetBstMidEndSepPunct{\mcitedefaultmidpunct}
{\mcitedefaultendpunct}{\mcitedefaultseppunct}\relax
\EndOfBibitem
\bibitem[Schaffer \latin{et~al.}(2012)Schaffer, Schaffer, and Ramasse]{schaffer2012sample}
Schaffer,~M.; Schaffer,~B.; Ramasse,~Q. Sample preparation for atomic-resolution STEM at low voltages by FIB. \emph{Ultramicroscopy} \textbf{2012}, \emph{114}, 62--71\relax
\mciteBstWouldAddEndPuncttrue
\mciteSetBstMidEndSepPunct{\mcitedefaultmidpunct}
{\mcitedefaultendpunct}{\mcitedefaultseppunct}\relax
\EndOfBibitem
\bibitem[Van~de Walle \latin{et~al.}(1999)Van~de Walle, Neugebauer, Stampfl, McCluskey, and Johnson]{van1999defects}
Van~de Walle,~C.~G.; Neugebauer,~J.; Stampfl,~C.; McCluskey,~M.; Johnson,~N. Defects and defect reactions in semiconductor nitrides. \emph{Acta Physica Polonica A} \textbf{1999}, \emph{96}, 613--627\relax
\mciteBstWouldAddEndPuncttrue
\mciteSetBstMidEndSepPunct{\mcitedefaultmidpunct}
{\mcitedefaultendpunct}{\mcitedefaultseppunct}\relax
\EndOfBibitem
\bibitem[Sebastian \latin{et~al.}(2017)Sebastian, Silva, and Sombra]{sebastian2017measurement}
Sebastian,~M.~T.; Silva,~M.; Sombra,~A. Measurement of microwave dielectric properties and factors affecting them. \emph{Microwave materials and applications 2V set} \textbf{2017}, 1--51\relax
\mciteBstWouldAddEndPuncttrue
\mciteSetBstMidEndSepPunct{\mcitedefaultmidpunct}
{\mcitedefaultendpunct}{\mcitedefaultseppunct}\relax
\EndOfBibitem
\bibitem[Akiyama \latin{et~al.}(2008)Akiyama, Kamohara, Kano, Teshigahara, and Kawahara]{akiyama2008influence}
Akiyama,~M.; Kamohara,~T.; Kano,~K.; Teshigahara,~A.; Kawahara,~N. Influence of oxygen concentration in sputtering gas on piezoelectric response of aluminum nitride thin films. \emph{Applied Physics Letters} \textbf{2008}, \emph{93}\relax
\mciteBstWouldAddEndPuncttrue
\mciteSetBstMidEndSepPunct{\mcitedefaultmidpunct}
{\mcitedefaultendpunct}{\mcitedefaultseppunct}\relax
\EndOfBibitem
\bibitem[Chen and Jaramillo(2017)Chen, and Jaramillo]{Jaramillo_Tauc}
Chen,~Z.; Jaramillo,~T.~F. The Use of UV-Visible Spectroscopy to Measure the Band Gap of a Semiconductor. \textbf{2017}, \relax
\mciteBstWouldAddEndPunctfalse
\mciteSetBstMidEndSepPunct{\mcitedefaultmidpunct}
{}{\mcitedefaultseppunct}\relax
\EndOfBibitem
\bibitem[Vachhani and Bhatnagar(2013)Vachhani, and Bhatnagar]{Vachhani}
Vachhani,~P.~S.; Bhatnagar,~A.~K. Oxygen pressure-dependent band gap in Cu-doped and -undoped ZnO films. \emph{Physica Scripta} \textbf{2013}, \emph{87}\relax
\mciteBstWouldAddEndPuncttrue
\mciteSetBstMidEndSepPunct{\mcitedefaultmidpunct}
{\mcitedefaultendpunct}{\mcitedefaultseppunct}\relax
\EndOfBibitem
\bibitem[Klein \latin{et~al.}(2023)Klein, Kampermann, Mockenhaupt, Behrens, Strunk, and Bacher]{Klein2023Tauc}
Klein,~J.; Kampermann,~L.; Mockenhaupt,~B.; Behrens,~M.; Strunk,~J.; Bacher,~G. Limitations of the Tauc Plot Method. \emph{Advanced Functional Materials} \textbf{2023}, \emph{33}\relax
\mciteBstWouldAddEndPuncttrue
\mciteSetBstMidEndSepPunct{\mcitedefaultmidpunct}
{\mcitedefaultendpunct}{\mcitedefaultseppunct}\relax
\EndOfBibitem
\end{mcitethebibliography}

% \begin{figure}
%   \includegraphics{Figures/TOC.png}
%   \caption*{}
% \end{figure}

\end{document}